\documentclass{article}
\usepackage{algorithm, algpseudocode, amsmath, amsthm, amssymb, rotating}

\newcommand{\abs}[1]{\left|#1\right|}

\newcommand{\ket}[1]{\left|#1\right\rangle}
\newcommand{\braket}[2]{\left\langle#1|#2\right\rangle}

\newcommand{\pref}[1]{(\ref{#1})}

\newtheorem{transformation-rule}{Transformation Rule}

\begin{document}

\title{Quantum Algorithms for Tree Isomorphism and State Symmetrization}
\author{David Rosenbaum \\ University of Washington, \\ Department of Computer Science and Engineering \\ Email: djr@cs.washington.edu}
\date{\today}
\maketitle

\begin{abstract}
The graph isomorphism problem is theoretically interesting and also has many practical applications.  The best known classical algorithms for graph isomorphism all run in time super-polynomial in the size of the graph in the worst case.  An interesting open problem is whether quantum computers can solve the graph isomorphism problem in polynomial time.  In this paper, an algorithm is shown which can decide if two rooted trees are isomorphic in polynomial time.  Although this problem is easy to solve efficiently on a classical computer, the techniques developed may be useful as a basis for quantum algorithms for deciding isomorphism of more interesting types of graphs.  The related problem of quantum state symmetrization is also studied.  A polynomial time algorithm for the problem of symmetrizing a set of orthonormal states over an arbitrary permutation group is shown.
\end{abstract}

\section{Introduction}
The problem of deciding if two graphs are isomorphic has many practical applications such as searching for an unknown molecule in a chemical database \cite{klin1997a}, verification of hierarchical circuits \cite{wong1985a} and generating application specific instruction sets \cite{cong2004a}.  It is not known to be solvable in polynomial time on a classical computer despite a great deal of effort to develop an efficient algorithm.  On the other hand, its \(\bf{NP}\)-completeness would imply the collapse of the polynomial hierarchy to the second level \cite{schoning1987a}.  Moreover, the complexity class \(\bf{NP} \cap \bf{coAM}\) that contains graph isomorphism \cite{babai1988a} is similar to \(\bf{NP} \cap \bf{coNP}\) which contains the decision version of factoring --- a problem that is solvable in quantum polynomial time using Shor's algorithm \cite{shor1994a}.  This suggests that graph isomorphism is of intermediate complexity and may be solvable in polynomial time on quantum computers.  This paper takes a step towards a polynomial time quantum algorithm for graph isomorphism based on the state symmetrization approach \cite{aharonov2003a} by developing quantum algorithms for rooted tree isomorphism.  By considering all possible roots in one of the trees, it is also possible to efficiently decide if two unrooted trees are isomorphic.  Although tree isomorphism can be decided in linear time on a classical computer \cite{aho1974a}, the quantum algorithm for tree isomorphism shown in this paper relies on a different strategy which may be useful for solving graph isomorphism in cases that are not known to be solvable on classical computers in polynomial time.  An algorithm is also presented for the related problem of symmetrizing a sequence of orthonormal states over an arbitrary permutation group.

\section{A quantum algorithm for tree isomorphism}
\label{tree-isomorphism-algorithm-section}
One approach to the graph isomorphism problem is to prepare a superposition of all possible permutations of the adjacency matrix \cite{aharonov2003a}.  For a graph \(G\) with automorphism group \(Aut(G)\) and adjacency matrix \(A\), this state is \(\ket{G} = \sqrt{\frac{Aut(G)}{n!}} \sum_{\pi \in S_n / Aut(G)} \ket{A^\pi}\) where \(A^\pi\) denotes the adjacency matrix obtained by applying \(\pi\) to \(A\).  Given two graphs \(G\) and \(H\), the states \(\ket{G}\) and \(\ket{H}\) are equal if \(G\) and \(H\) are isomorphic and are orthogonal otherwise.  It can then be determined if \(G\) and \(H\) are isomorphic using the swap test \cite{buhrman2001a}.  However, this is not sufficient to solve graph isomorphism since no efficient algorithm is known for preparing the state \(\ket{G}\).  As a first step towards developing a general quantum algorithm for graph isomorphism using this approach, a quantum algorithm for tree isomorphism based on state symmetrization will be presented.  The tree isomorphism algorithm operates on qudits with four computational basis states rather than qubits.  These states are \(\ket{0}\), \(\ket{1}\), \(\ket{\alpha}\) and \(\ket{\beta}\).  The states \(\ket{0}\) and \(\ket{1}\) are used in most parts of the algorithm for representing binary numbers and permutations while the other computational basis states have special purposes. The state \(\ket{\alpha}\) is used to mark the beginning of a tree while \(\ket{\beta}\) marks the end of a tree as in the classical tree isomorphism algorithm.  The algorithm works by preparing a state \(\ket{T}\) for a given rooted tree \(T\) such that given two rooted trees \(T_1\) and \(T_2\), if \(\ket{T_1}\) and \(\ket{T_2}\) are of the same dimension then \(\braket{T_1}{T_2} = 1\) if and only if \(T_1 \cong T_2\) and \(\braket{T_1}{T_2} = 0\) if and only if \(T_1 \not\cong T_2\).  If \(\ket{T_1}\) and \(\ket{T_2}\) are not of the same dimension then it is required that \(T_1 \not\cong T_2\).  Such states can be prepared recursively as follows.  Consider a rooted tree \(T\).  If \(T\) consists of a single node (for instance if \(T\) is a leaf node in a larger tree), then let \(\ket{T} = \ket{\alpha, \beta}\).  Otherwise, let \(T_1, \ldots, T_k\) denote the subtrees rooted at the \(k\) children of the root node of \(T\).  For each \(1 \leq i \leq k\), let \(U_i\) denote the unitary matrix such that \(U_i \ket{0} = \ket{T_i}\).  Note that the states \(\ket{T_i}\) and \(\ket{T_j}\) are not necessarily of the same dimension for \(T_i \not\cong T_j\) as there are bijections between the number of qudits in \(\ket{T_i}\), the dimension of \(\ket{T_i}\) and the number of nodes in the tree \(T_i\).  Let \(b_i\) denote the number of qudits in the state \(\ket{T_i}\).  Observe that (assuming the algorithm works properly on smaller trees) \(b_i \not= b_j\) or \(\braket{T_i}{T_j} = 0\) if \(T_i \not\cong T_j\) while \(b_i = b_j\) and \(\braket{T_i}{T_j} = 1\) if \(T_i \cong T_j\).  Now construct the state
\begin{equation}
  \bigotimes_{i = 1}^k \ket{T_i}_{T_i} \otimes \ket{0}_{M_i}
\end{equation}
where each register \(T_i\) contains \(b_i\) qudits and each register \(M_i\) contains \(\left\lceil \log_2 \ell \right\rceil\) qudits where \(\ell\) is the largest branching factor that occurs in the entire tree.  Let \(c_i\) be the number of subtrees \(T_j\) such that \(n_i = n_j\) where \(n_i\) is the number of nodes in \(T_i\).  Now execute algorithm \ref{isomorphic-subtree-counting-algorithm} to obtain the state
\begin{equation}
  \bigotimes_{i = 1}^k \ket{T_i}_{T_i} \otimes \ket{m(i)}_{M_i}
\end{equation}
where \(m(i)\) is the number of subtrees \(T_j\) with \(j < i\) that are isomorphic to \(T_i\).
\begin{algorithm}
  \begin{center}
    \begin{algorithmic}[1]
      \For{\(i = 1, \ldots, k\)}
        \If{\(c_i > 1\)}
          \For{\(j = 1, \ldots, i - 1\)}
            \If{\(b_i = b_j\)}
              \State Apply \(U_i^\dagger\) to register \(T_j\) resulting in the state \(\ket{0}\) if and only if \(T_i \cong T_j\)
              \State Add one to the contents of the register \(M_i\) conditional on the contents of register \(T_j\) being \(\ket{0}\)
              \State Apply \(U_i\) to register \(T_j\) in order to restore the original state
            \EndIf
          \EndFor
        \EndIf
      \EndFor
    \end{algorithmic}
  \end{center}
  \caption{Counting the number of subtrees \(T_j\) such that \(j < i\) and \(T_i \cong T_j\) for each \(1 \leq i \leq k\)}
  \label{isomorphic-subtree-counting-algorithm}
\end{algorithm}
Now append a new set of registers initialized to \(\ket{0}\) yielding the state
\begin{equation}
  \bigotimes_{i = 1}^k \ket{T_i}_{T_i} \otimes \ket{m(i)}_{M_i} \bigotimes_{i = 1}^k \ket{0}_{N_i}
\end{equation}
where each register \(N_i\) uses \(\left \lceil \log_2 \ell \right \rceil\) qudits.  XOR the contents of each register \(M_i\) into the register \(N_i\) obtaining the state
\begin{equation}
  \bigotimes_{i = 1}^k \ket{T_i}_{T_i} \otimes \ket{m(i)}_{M_i} \bigotimes_{i = 1}^k \ket{m(i)}_{N_i}
\end{equation}
Letting \(\ket{\tau_i} = \ket{T_i} \otimes \ket{m(i)}\) results in the state
\begin{equation}
  \bigotimes_{i = 1}^k \ket{\tau_i}_{T_i^\prime} \bigotimes_{i = 1}^k \ket{m(i)}_{N_i}
\end{equation}
where each register \(T_i^\prime\) contains registers \(T_i\) and \(M_i\).  Now, prepare the state \(\ket{S_k}\) (the uniform superposition over all permutations in \(S_k\)) in another register to obtain
\begin{equation}
  \ket{S_k}_S \bigotimes_{i = 1}^k \ket{\tau_i}_{T_i^\prime} \bigotimes_{i = 1}^k \ket{m(i)}_{N_i}
\end{equation}
Apply a conditional permutation which permutes the order of the registers \(T_i^{\prime}\) resulting in the state
\begin{equation}
  \frac{1}{\sqrt{k!}} \sum_{\pi \in S_k} \ket{\pi}_S \bigotimes_{i = 1}^k \ket{\tau_{\pi^{-1}(i)}}_{T_i^\prime} \bigotimes_{i = 1}^k \ket{m(i)}_{N_i}
\end{equation}
Note that since the registers \(T_i^\prime\) now contain different numbers of qudits for different terms in the quantum superposition, special care must be taken here.  Each \(\pi \in S_k\) is represented as a product \(g_k \cdots g_2\) where each \(g_i \in \left\{(1, i), \ldots, (i - 1, i), \iota\right\}\) and the permutation \(\ket{\pi}\) is stored as \(\ket{g_k} \otimes \cdots \otimes \ket{g_2}\) where \(\ket{g_i} = \ket{j}\) when \(g_i = (j, i)\).  The next step is to uncompute the contents of the \(S\) register.  Each vector \(\bigotimes_{i=1}^k \ket{\tau_{\pi^{-1}(i)}}_{T_i^\prime}\) corresponds to a unique permutation because each \(\ket{\tau_i} = \ket{T_i} \otimes \ket{m(i)}\) is orthogonal to every \(\ket{\tau_j} = \ket{T_j} \otimes \ket{m(j)}\) for \(i \not= j\).  Note that \(T_i\) might be isomorphic to \(T_j\) for some \(i \not= j\) but in this case \(m(i) \not= m(j)\) so that \(\ket{\tau_i}\) and \(\ket{\tau_j}\)  will still be orthogonal.  This was the reason for labeling each register \(T_i\) with the number of isomorphic subtrees in the sequence \(T_1, \ldots, T_{i - 1}\).  Add another set of registers initialized to \(\ket{0}\) obtaining the state
\begin{equation}
  \frac{1}{\sqrt{k!}} \sum_{\pi \in S_k} \bigotimes_{i = 1}^k \ket{0}_{P_i} \otimes \ket{\pi}_S \bigotimes_{i = 1}^k \ket{\tau_{\pi^{-1}(i)}}_{T_i^\prime} \bigotimes_{i = 1}^k \ket{m(i)}_{N_i}
\end{equation}
where each register \(P_i\) uses \(\left\lceil \log_2(k + 1) \right\rceil\) qudits.  Now XOR the contents of register \(P_i\) with \(\ket{i}\) resulting in the state
\begin{equation}
  \frac{1}{\sqrt{k!}} \sum_{\pi \in S_k} \bigotimes_{i = 1}^k \ket{i}_{P_i} \otimes \ket{\pi}_S \bigotimes_{i = 1}^k \ket{\tau_{\pi^{-1}(i)}}_{T_i^\prime} \bigotimes_{i = 1}^k \ket{m(i)}_{N_i}
\end{equation}
Apply a conditional permutation to reorder the registers \(P_1, \ldots, P_k\) according to the contents of register \(S\) obtaining the state
\begin{equation}
  \frac{1}{\sqrt{k!}} \sum_{\pi \in S_k} \bigotimes_{i = 1}^k \ket{\pi^{-1}(i)}_{P_i} \otimes \ket{\pi}_S \bigotimes_{i = 1}^k \ket{\tau_{\pi^{-1}(i)}}_{T_i^\prime} \bigotimes_{i = 1}^k \ket{m(i)}_{N_i}
\end{equation}
Now, note that each \(\ket{\pi}\) is represented in subgroup tower form as a product \(g_k \cdots g_2\) where each \(g_i \in \left\{(1, i), \ldots, (i - 1, i), \iota\right\}\).  Then the contents of register \(S\) can be uncomputed using the procedure (to be described later) for converting the row representation \(\bigotimes_{i = 1}^k \ket{\pi^{-1}(i)}_{P_i}\) of \(\pi^{-1}\) to the subgroup tower representation \(\ket{\pi}\) for \(\pi\).  This results in the state
\begin{equation}
  \frac{1}{\sqrt{k!}} \sum_{\pi \in S_k} \bigotimes_{i = 1}^k \ket{\pi^{-1}(i)}_{P_i} \bigotimes_{i = 1}^k \ket{\tau_{\pi^{-1}(i)}}_{T_i^\prime} \bigotimes_{i = 1}^k \ket{m(i)}_{N_i}
\end{equation}
after discarding the contents of register \(S\) which is now \(\ket{0}\).  The next step is to uncompute each \(P_i\) register.  To do this, first add a new register initialized to \(\ket{0}\) yielding the state
\begin{equation}
  \frac{1}{\sqrt{k!}} \sum_{\pi \in S_k} \ket{0}_B \bigotimes_{i = 1}^k \ket{\pi^{-1}(i)}_{P_i} \bigotimes_{i = 1}^k \ket{\tau_{\pi^{-1}(i)}}_{T_i^\prime} \bigotimes_{i = 1}^k \ket{m(i)}_{N_i}
\end{equation}
where the register \(B\) uses one qudit.  The contents of the \(P_i\) registers are then uncomputed using algorithm \ref{row-form-uncomputation-algorithm}.  This results in the state
\begin{equation}
  \frac{1}{\sqrt{k!}} \sum_{\pi \in S_k} \bigotimes_{i = 1}^k \ket{\tau_{\pi^{-1}(i)}}_{T_i^\prime} \bigotimes_{i = 1}^k \ket{m(i)}_{N_i}
\end{equation}
after discarding the contents of the \(P_i\) and \(B\) registers which are now all in the state \(\ket{0}\).
\begin{algorithm}
  \begin{center}
    \begin{algorithmic}[1]
      \For{\(i = 1, \ldots, k\)}
        \For{\(j = 1, \ldots, k\)}
          \State XOR the contents of register \(B\) with \(1\) conditional on register \(T_j\) containing \(b_i\) qudits
          \If{\(c_i > 1\)}
            \State Apply \(U_i^\dagger\) to register \(T_j\) conditional on the contents of register \(B\) being \(\ket{1}\)
            \State XOR the contents of register \(P_j\) with \(i\) conditional on the contents of register \(B\) being \(\ket{1}\), the contents of register \(T_j\) being \(\ket{0}\) and the contents of register \(M_j\) being equal to register \(N_i\)
            \State Apply \(U_i\) to register \(T_j\) conditional on the contents of register \(B\) being \(\ket{1}\) to restore the register to its original state
          \Else
            \State XOR the contents of register \(P_j\) with \(i\) conditional on the contents of register \(B\) being \(\ket{1}\)
          \EndIf
          \State XOR the contents of register \(B\) with \(1\) conditional on register \(T_j\) containing \(b_i\) qudits, restoring register \(B\) to the state \(\ket{0}\)
        \EndFor
      \EndFor
    \end{algorithmic}
  \end{center}
  \caption{Uncomputing \(P_i\) for all \(1 \leq i \leq k\)}
  \label{row-form-uncomputation-algorithm}
\end{algorithm}
It is now necessary to uncompute the contents of the \(N_i\) registers.  First add a new set of registers initialized to \(\ket{0}\) yielding the state
\begin{equation}
  \frac{1}{\sqrt{k!}} \sum_{\pi \in S_k} \bigotimes_{i = 1}^k \ket{\tau_{\pi^{-1}(i)}}_{T_i^\prime} \bigotimes_{i = 1}^k \ket{0}_{T_i^{\prime \prime}} \bigotimes_{i = 1}^k \ket{m(i)}_{N_i}
\end{equation}
where each register \(T_i^{\prime \prime}\) uses \(b_i\) qudits.  Initialize the contents of each register \(T_i^{\prime \prime}\) where \(c_i > 1\) to \(\ket{T_i}\) by applying \(U_i\) to each register \(T_i\) where \(c_i > 1\).  This results in the state
\begin{equation}
  \frac{1}{\sqrt{k!}} \sum_{\pi \in S_k} \bigotimes_{i = 1}^k \ket{\tau_{\pi^{-1}(i)}}_{T_i^\prime} \bigotimes_{i = 1}^k \ket{T_i^\prime}_{T_i^{\prime \prime}} \bigotimes_{i = 1}^k \ket{m(i)}_{N_i}
\end{equation}
where \(\ket{T_i^\prime} = \ket{T_i}\) if \(c_i > 1\) and \(\ket{T_i^\prime} = \ket{0}\) if \(c_i = 1\).  The contents of the \(N_i\) registeres are then uncomputed by applying the inverse of algorithm \ref{isomorphic-subtree-counting-algorithm} which was used to set each register \(N_i\) to \(\ket{m(i)}\) in the first place.  This yields the state 
\begin{equation}
  \frac{1}{\sqrt{k!}} \sum_{\pi \in S_k} \bigotimes_{i = 1}^k \ket{\tau_{\pi^{-1}(i)}}_{T_i^\prime} \bigotimes_{i = 1}^k \ket{T_i^\prime}_{T_i^{\prime \prime}}
\end{equation}
after the \(N_i\) registers which are now all equal to \(\ket{0}\) are discarded.  Uncomputing the contents of the \(T_i^{\prime \prime}\) registers is accomplished by applying \(U_i^\dagger\) to each register \(T_i^{\prime \prime}\) where \(c_i > 1\).  Discarding the registers \(T_i^{\prime \prime}\) then results in the state
\begin{equation}
  \frac{1}{\sqrt{k!}} \sum_{\pi \in S_k} \bigotimes_{i = 1}^k \ket{\tau_{\pi^{-1}(i)}}_{T_i^\prime}
\end{equation}
Prepending a \(\ket{\alpha}\) and appending a \(\ket{\beta}\) to this state yields the final state
\begin{equation}
  \ket{T} = \frac{1}{\sqrt{k!}} \ket{\alpha} \otimes \left(\sum_{\pi \in S_k} \bigotimes_{i = 1}^k \ket{\tau_{\pi^{-1}(i)}}_{T_i^\prime}\right) \otimes \ket{\beta}
\end{equation}
Given two rooted trees \(T_1\) and \(T_2\), the states \(\ket{T_1}\) and \(\ket{T_2}\) as defined in the above equation have the property that if \(\ket{T_1}\) and \(\ket{T_2}\) are of the same dimension then \(\braket{T_1}{T_2} = 1\) if and only if \(T_1 \cong T_2\) and \(\braket{T_1}{T_2} = 0\) if and only if \(T_1 \not\cong T_2\) as desired.

\section{Complexity analysis of the tree isomorphism algorithm}
In this section, it will be shown that the quantum tree isomorphism algorithm presented in section \ref{tree-isomorphism-algorithm-section} is polynomial.  To do this, the first step is to create a recurrence for the running time of the algorithm.  This results in
\begin{equation}
  T(n) = \sup_{\begin{array}{c}
                   \sum_{i = 1}^k n_i = n - 1 \\
                   1 \leq k \leq n -1 \\
                   1 \leq n_i, n_i \in \mathbb{N}
                 \end{array}} \left \lbrace t_{n_1, \ldots, n_k} \right \rbrace
\end{equation}
where
\begin{align}
  t_{n_1, \ldots, n_k} = & \sum_{i = 1}^k T(n_i) \nonumber \\
                    & {} + \sum_{i = 1}^k [c_i > 1] \sum_{j = 1}^{i - 1} [n_i = n_j] \left (2 T(n_i) + O(\log \ell) \right) \nonumber \\
                    & {} + O(k \log \ell) \nonumber \\
                    & {} + O(k^2 \log \ell) \nonumber \\
                    & {} + p_1(S(n), k) \nonumber \\
                    & {} + O(k \log k) \nonumber \\
                    & {} + p_1(k \left \lceil \log_2 (k + 1) \right \rceil, k) \nonumber \\
                    & {} + p_2(k) \nonumber \\
                    & {} + \sum_{i = 1}^k \sum_{j = 1}^k \left(2 p_3(n, k, b_i) + [c_i > 1] \left (2 T(n_i) + O(\log \ell) \right) + [c_i = 1] O(\log k) \right ) \nonumber \\
                    & {} + \sum_{i = 1}^k [c_i > 1] T(n_i) \nonumber \\
                    & {} + \sum_{i = 1}^k [c_i > 1] \sum_{j = 1}^{i - 1} [n_i = n_j] \left (2 T(n_i) + O(\log \ell) \right) \nonumber \\
                    & {} + \sum_{i = 1}^k [c_i > 1] T(n_i) \nonumber \\
                    & {} + O(1)
\end{align}
and
\begin{equation}
  S(n) = 2 n + \left \lceil \log_2 \ell \right \rceil (n - 1)
\end{equation}
is the number of qudits required for the state \(\ket{T}\) where \(T\) is a tree containing \(n\) nodes.  Each \(c_i\) denotes the multiplicity with which \(n_i\) occurs in the values \(n_j\) as in the description of the algorithm.  The functions \(p_i\) describe the complexities of different subroutines of the algorithm and are bounded above by polynomials.  The recurrence \(T(n)\) is bounded above by the recurrence
\begin{equation}
  T^\prime(n) = \sup_{\begin{array}{c}
                   \sum_{i = 1}^k n_i = n - 1 \\
                   1 \leq k \leq n -1 \\
                   1 \leq n_i, n_i \in \mathbb{N}
                 \end{array}} \left \lbrace t_{n_1, \ldots, n_k}^\prime \right \rbrace
\end{equation}
where
\begin{equation}
  t_{n_1, \ldots, n_k}^\prime = \sum_{i = 1}^k \left(1 + 2 [c_i > 1] + 4 [c_i > 1] c_i + 2 k [c_i > 1] \right) T^\prime(n_i) + p(n, \ell)
\end{equation}
and \(p\) is bounded above by some polynomial.  It will now be shown that \(T^\prime(n)\) is polynomial in \(n\).  Consider \(T^\prime(n + 1)\).  There exist \(k\) and \(n_i\) with \(1 \leq k \leq n\) and \(\sum_{i = 1}^k n_i = n\) such that \(T^\prime(n + 1) = \sum_{i = 1}^k \left(1 + 2 [c_i > 1] + 4 [c_i > 1] c_i + 2 k [c_i > 1] \right) T^\prime(n_i) + p(n + 1, \ell)\).  Let \(C\) be a subset of \(\left \lbrace 1, \ldots, k \right \rbrace\) such that for each \(n_i\), there exists \(j \in C\) such that \(n_i = n_j\) and if \(i \not= j \in C\) then \(n_i \not= n_j\).  Thus, for each distinct value of \(n_i\), \(C\) contains exactly one index \(j\) such that \(n_i = n_j\).  The analysis is performed using \(T^\prime(n + 1) - T^\prime(n)\).  Consider the following three cases:
\begin{itemize}
\item
  Suppose that \(n_j = 1\) for some \(1 \leq j \leq k\) with \(j \in C\).  One of the subtrees with only one node will now be removed to obtain a lower bound for \(T^\prime(n)\).  Let

  \begin{equation}
    c_i^\prime = \begin{cases}
                 c_i - 1 & n_i = 1 \\
                 c_i & n_i > 1
               \end{cases}
  \end{equation}
  Then a lower bound for \(T^\prime(n)\) is

  \begin{equation}
    T^\prime(n) \geq \sum_{\begin{array}{c} 1 \leq i \leq k \\ i \not= j\end{array}} \left(1 + 2 [c_i^\prime > 1] + 4 [c_i^\prime > 1] c_i^\prime + 2 k [c_i^\prime > 1] \right) T^\prime(n_i) + p(n, \ell)
  \end{equation}
  Therefore,

  \begin{align}
    T^\prime(n + 1) - T^\prime(n) \leq & \sum_{1 \leq i \leq k} \left(1 + 2 [c_i > 1] + 4 [c_i > 1] c_i + 2 k [c_i > 1] \right) T^\prime(n_i) + p(n + 1, \ell) \nonumber \\
                                    & {} - \sum_{\begin{array}{c} 1 \leq i \leq k \\ i \not= j\end{array}} \left(1 + 2 [c_i^\prime > 1] + 4 [c_i^\prime > 1] c_i^\prime + 2 k [c_i^\prime > 1] \right) T^\prime(n_i) - p(n, \ell) \\
                                  = & \sum_{i \in C} c_i \left(1 + 2 [c_i > 1] + 4 [c_i > 1] c_i + 2 k [c_i > 1] \right) T^\prime(n_i) + p(n + 1, \ell) \nonumber \\
                                    & {} - \sum_{i \in C} c_i^\prime \left(1 + 2 [c_i^\prime > 1] + 4 [c_i^\prime > 1] c_i^\prime + 2 k [c_i^\prime > 1] \right) T^\prime(n_i) - p(n, \ell) \\
                               \leq & \sum_{i \in C} c_i \left(1 + 2 [c_i > 1] + 4 [c_i > 1] c_i + 2 k [c_i > 1] \right) T^\prime(n_i) + p(n + 1, \ell) \nonumber \\
                                    & {} - \sum_{i \in C \setminus \lbrace j \rbrace} c_i \left(1 + 2 [c_i > 1] + 4 [c_i > 1] c_i + 2 k [c_i > 1] \right) T^\prime(n_i) - p(n, \ell) \\
                                  = & c_j \left(1 + 2 [c_j > 1] + 4 [c_j > 1] c_j + 2 k [c_j > 1] \right) T^\prime(n_j) + p(n + 1, \ell) - p(n, \ell) \\
                               \leq & n (3 + 6 n) T^\prime(1) + p(n + 1, \ell) - p(n, \ell)
  \end{align}
\item
  Consider the case where \(c_j = 1\) for some \(1 \leq j \leq k\).  A lower bound for \(T^\prime(n)\) is now obtained by removing one node from the subtree containing \(n_j\) nodes.  Let

  \begin{equation}
    n_i^\prime = \begin{cases}
                 n_j - 1 & i = j \\
                 n_i & i \not= j
               \end{cases}
  \end{equation}
  and let \(c_i^\prime\) denote the multiplicity with which \(n_i^\prime\) occurs in the values \(n_1^\prime, \ldots, n_k^\prime\).  Then a lower bound for \(T^\prime(n)\) is

  \begin{equation}
    T^\prime(n) \geq \sum_{i = 1}^k \left(1 + 2 [c_i^\prime > 1] + 4 [c_i^\prime > 1] c_i^\prime + 2 k [c_i^\prime > 1] \right) T^\prime(n_i^\prime) + p(n, \ell)
  \end{equation}
  
  Thus,

  \begin{align}
    T^\prime(n + 1) - T^\prime(n) \leq & \sum_{i = 1}^k \left(1 + 2 [c_i > 1] + 4 [c_i > 1] c_i + 2 k [c_i > 1] \right) T^\prime(n_i) + p(n + 1, \ell) \nonumber \\
                                    & {} - \sum_{i = 1}^k \left(1 + 2 [c_i^\prime > 1] + 4 [c_i^\prime > 1] c_i^\prime + 2 k [c_i^\prime > 1] \right) T^\prime(n_i^\prime) - p(n, \ell) \\
                               \leq & \sum_{i = 1}^k \left(1 + 2 [c_i > 1] + 4 [c_i > 1] c_i + 2 k [c_i > 1] \right) T^\prime(n_i) + p(n + 1, \ell) \nonumber \\
                                    & {} - \sum_{i = 1}^k \left(1 + 2 [c_i > 1] + 4 [c_i > 1] c_i + 2 k [c_i > 1] \right) T^\prime(n_i^\prime) - p(n, \ell) \\
                                  = & T^\prime(n_j) - T^\prime(n_j^\prime) + p(n + 1, \ell) - p(n, \ell) \\
                                  = & T^\prime(n_j) - T^\prime(n_j - 1) + p(n + 1, \ell) - p(n, \ell) \\
                               \leq & \sup_{\begin{array}{c} 1 \leq n^\prime \leq n \\ n^\prime \in \mathbb{N}\end{array}} \left \lbrace T^\prime(n^\prime) - T^\prime(n^\prime - 1) \right \rbrace + p(n + 1, \ell) - p(n, \ell)
  \end{align}
\item
  For the final case, suppose that \(c_i > 1\) and \(n_i > 1\) for all \(1 \leq i \leq k\).  Choose the smallest \(n_j\) such that \(j \in C\).  A lower bound for \(T^\prime(n)\) will be obtained by reducing each \(n_i\) such that \(n_i = n_j\) by one.  Let

  \begin{equation}
    n_i^\prime = \begin{cases}
                 n_i - 1 & n_i = n_j \\
                 n_i & n_i \not= n_j
               \end{cases}
  \end{equation}
  Then a lower bound for \(T^\prime(n)\) is

  \begin{equation}
    T^\prime(n) \geq \sum_{i = 1}^k \left(1 + 2 [c_i > 1] + 4 [c_i > 1] c_i + 2 k [c_i > 1] \right) T^\prime(n_i^\prime) + p(n, \ell)
  \end{equation}
  Hence,

  \begin{align}
    T^\prime(n + 1) - T^\prime(n) \leq & \sum_{i = 1}^k \left(1 + 2 [c_i > 1] + 4 [c_i > 1] c_i + 2 k [c_i > 1] \right) T^\prime(n_i) + p(n + 1, \ell) \nonumber \\
                                       & {} - \sum_{i = 1}^k \left(1 + 2 [c_i > 1] + 4 [c_i > 1] c_i + 2 k [c_i > 1] \right) T^\prime(n_i^\prime) - p(n, \ell) \\
                                     = & \sum_{i \in C} c_i \left(1 + 2 [c_i > 1] + 4 [c_i > 1] c_i + 2 k [c_i > 1] \right) T^\prime(n_i) + p(n + 1, \ell) \nonumber \\
                                       & {} - \sum_{i \in C} c_i \left(1 + 2 [c_i > 1] + 4 [c_i > 1] c_i + 2 k [c_i > 1] \right) T^\prime(n_i^\prime) - p(n, \ell) \\
                                     = & \sum_{i \in C} c_i \left(3 + 4 c_i + 2 k \right) T^\prime(n_i) + p(n + 1, \ell) \nonumber \\
                                       & {} - \sum_{i \in C} c_i \left(3 + 4 c_i + 2 k \right) T^\prime(n_i^\prime) - p(n, \ell) \\
                                     = & c_j \left(3 + 4 c_j + 2 k \right) \left(T^\prime(n_j) - T^\prime(n_j^\prime) \right) + p(n + 1, \ell) - p(n, \ell) \\
                                     = & c_j \left(3 + 4 c_j + 2 k \right) \left(T^\prime(n_j) - T^\prime(n_j - 1) \right) + p(n + 1, \ell) - p(n, \ell) \\
                                  \leq & \sup_{\begin{array}{c} k \leq k^\prime \leq n - 1 \\ k^\prime \in \mathbb{Q}\end{array}} \left \lbrace k (3 + 6 k) \left (T^\prime \left(\left \lfloor\frac{n}{k^\prime}\right \rfloor \right) - T^\prime \left(\left \lfloor\frac{n}{k^\prime}\right \rfloor - 1 \right) \right) \right \rbrace \nonumber \\
                                       & {} + p(n + 1, \ell) - p(n, \ell)
  \end{align}
\end{itemize}
A recurrence \(D(n)\) can now be defined in terms of maximum of the upper bounds obtained in each case so that
\begin{equation}
  T^\prime(n + 1) - T^\prime(n) \leq D(n + 1)
\end{equation}
Let
\begin{equation}
  D(n + 1) = \max \left \lbrace f(n), g(n), h(n) \right \rbrace + d(n + 1, \ell)
\end{equation}
where
\begin{align}
  f(n) = & n (3 + 6 n) T^\prime(1) \\
  g(n) = & \sup_{\begin{array}{c} 1 \leq n^\prime \leq n \\ n^\prime \in \mathbb{N}\end{array}} \left \lbrace D(n^\prime) \right \rbrace \\
  h(n) = & \sup_{\begin{array}{c} k \leq k^\prime \leq n - 1 \\ k^\prime \in \mathbb{Q}\end{array}} \left \lbrace k (3 + 6 k) D \left(\left \lfloor\frac{n}{k^\prime}\right \rfloor \right) \right \rbrace
\end{align}
and \(d(n + 1, \ell)\) is a polynomial upper bound for \(p(n + 1, \ell) - p(n, \ell)\).  Also, note \(k \geq 2\) is a value which depends on \(n\).  It will now be shown that \(D(n) \leq a n^p \ell^q\) for all \(n\) and for all \(\ell\) with appropriately chosen constants \(a\), \(p\) and \(q\).  Let \(\deg_n d\) and \(\deg_\ell d\) denote the degrees of \(d\) in \(n\) and \(\ell\) respectively.  Let \(p = \max \lbrace \deg_n d + 1, 6 \rbrace\) and \(q = \deg_\ell d\).  Since \((n + 1)^p = \sum_{k = 0}^p \binom{p}{k} n^k\), \((n + 1)^p - n^p \geq p n^{p - 1}\).  This implies that by choosing a constant \(a > 0\) which depends only on the coefficients of \(d(n ,\ell)\) and the constants \(D(1)\) and \(T^\prime(1)\), the following properties can be satisfied:
\begin{itemize}
\item
  \(D(1) \leq a\)
\item
  \(f(n) \leq a n^p \ell^q\) for all \(n\) and for all \(\ell\)
\item
  \(d(n + 1, \ell) \leq a(n + 1)^p \ell^q - a n^p \ell^q\) for all \(n\) and for all \(\ell\)
\end{itemize}
It can then be shown that \(D(n) \leq a n^p \ell^q\) for all \(n\) and for all \(\ell\) by induction.  Note that the basis case is satisfied by the choice of \(a\).  For the inductive case, assume that \(D(n^\prime) \leq a {n^\prime}^p \ell^q\) for \(n^\prime \leq n\).  Now consider \(D(n + 1)\).  By choice of \(a\), \(f(n) \leq a n^p \ell^q\) so
\begin{equation}
  f(n) + d(n + 1, \ell) \leq a (n + 1)^p \ell^q
\end{equation}
By assumption, \(D(n^\prime) \leq a {n^\prime}^p \ell^q\) so for \(1 \leq n^\prime \leq n\), \(D(n^\prime) \leq a n^p \ell^q\).  Thus,
\begin{equation}
  g(n) + d(n + 1, \ell) \leq a (n + 1)^p \ell^q
\end{equation}
Recall that in the expression for \(h(n)\), \(k \geq 2\).  Therefore, for \(k \leq k^\prime \leq n - 1\) with \(k^\prime \in \mathbb{Q}\),
\begin{align}
  k (3 + 6 k) D \left(\left \lfloor\frac{n}{k^\prime}\right \rfloor \right) \leq & a k (3 + 6 k) \left \lfloor\frac{n}{k^\prime}\right \rfloor^p \ell^q \\
                                                                           \leq & a k (3 + 6 k) \left(\frac{n}{k^\prime}\right)^p \ell^q \\
                                                                           \leq & 9 a k^2 \left(\frac{n}{k^\prime}\right)^p \ell^q \\
                                                                           \leq & a n^p \ell^q
\end{align}
Thus, \(h(n) \leq a n^p \ell^q\) so
\begin{equation}
  h(n) + d(n + 1, \ell) \leq a (n + 1)^p \ell^q
\end{equation}
Consequently,
\begin{equation}
  D(n + 1) \leq a (n + 1)^p \ell^q
\end{equation}
which proves that
\begin{equation}
  D(n) \leq a n^p \ell^q
\end{equation}
for all \(n\) and for all \(\ell\).  Hence, 
\begin{align}
  T(n) \leq & T^\prime(n) \\
          = & \sum_{i = 1}^n (T^\prime(i) - T^\prime(i - 1)) \\
       \leq & \sum_{i = 1}^n D(i) \\
       \leq & a \ell^q \sum_{i = 1}^n  i^p \\
       \leq & a \sum_{i = 1}^n  n^p \ell^q \\
          = & a n^{p + 1} \ell^q \\
       \leq & a n^{p + q + 1} \\
          = & O(n^{p + q + 1})
\end{align}
so \(T(n)\) is polynomial in \(n\).

\section{Subroutines for the tree isomorphism algorithm}
To finish describing the algorithm, all that remains is to show how to implement the three subroutines it utilizes.  These are preparing the uniform superposition over all permutations, applying a conditional permutation and converting the row form \(\bigotimes_{i = 1}^k \ket{\pi^{-1}(i)}\) of the inverse of a permutation \(\pi \in S_k\) to the subgroup tower form \(\ket{\pi}\) if \(\pi\).

\subsection{Preparing the uniform superposition over all permutations}
This state can be prepared for \(S_k\) using \(O(k^2 \log k)\) basic operations using generator state based algorithms for initialization \cite{ventura1999a,rosenbaum2008a,rosenbaum2009a}.  It is actually possibly to do this more efficiently but this is not necessary for the purposes of this algorithm as it will result in the same complexity.

\subsection{Implementing conditional permutations}
Since this operation is a bijection which takes an input of fixed length that can be implemented in polynomial time on a classical computer it can also be implemented in polynomial time on a quantum computer.

\subsection{Converting permutations from row form to subgroup tower form}
This section will show how to convert a permutation \(\pi \in S_k\) in row form
\begin{equation}
  \bigotimes_{i = 1}^k \ket{\pi(i)}_{P_i}
\end{equation}
to the subgroup tower form \(\ket{\pi}\).  First, observe that \((\pi(k), k) \pi \in S_{k - 1}\) so the equation
\begin{equation}
  \pi = (\pi(k), k) \left((\pi(k), k) \pi\right)
\end{equation}
allows \(\pi\) to be decomposed into the product of one of the transpositions in the set \(\left\{(1, k), \ldots, (k - 1, k), \iota\right\}\) and an element of \(S_{k - 1}\).  The conversion procedure works by repeating this decomposition.  Append a second set of registers initialized to \(\ket{0}\) resulting in the state
\begin{equation}
  \bigotimes_{i = 1}^k \ket{\pi(i)}_{P_i} \bigotimes_{i = k}^2 \ket{0}_{G_i}
\end{equation}
where the register \(G_i\) contains \(\left\lceil \log_2(i + 1) \right\rceil\) qudits.  Now execute algorithm \ref{row-form-to-subgroup-tower-form-conversion-algorithm}.  This results in the state
\begin{equation}
  \bigotimes_{i = 1}^k \ket{i}_{P_i} \bigotimes_{i = k}^2 \ket{g_k}_{G_i}
\end{equation}
\begin{algorithm}
  \begin{center}
    \begin{algorithmic}[1]
      \For{\(i = k, \ldots, 2\)}
        \State Copy the contents of register \(P_i\) to register \(G_i\)
        \For{\(j = 1, \ldots, k\)}
          \State Swap the contents of registers \(P_i\) and \(P_j\) conditional on \(j\) being equal to the contents of register \(G_i\)
        \EndFor
      \EndFor
    \end{algorithmic}
  \end{center}
  \caption{Converting a permutation in row form to subgroup tower form}
  \label{row-form-to-subgroup-tower-form-conversion-algorithm}
\end{algorithm}
where \(\pi = g_k \cdots g_2\) is the subgroup tower form of \(\pi\).  Discarding the registers \(P_i\) yields
\begin{equation}
  \ket{\pi} = \bigotimes_{i = k}^2 \ket{g_k}_{G_i}
\end{equation}
which is the subgroup tower form of \(\pi\).  Now the algorithm actually requires a procedure that will convert the row form
\begin{equation}
  \bigotimes_{i = 1}^k \ket{\pi^{-1}(i)}_{P_i}
\end{equation}
of \(\pi^{-1}\) to the subgroup tower form \(\ket{\pi}\) of \(\pi\).  To do this it suffices to show how to invert the row form
\begin{equation}
  \bigotimes_{i = 1}^k \ket{\pi(i)}_{P_i}
\end{equation}
of \(\pi\).  This can be done by first appending new registers each in the state \(\ket{0}\).  This results in the state
\begin{equation}
  \bigotimes_{i = 1}^k \ket{\pi(i)}_{P_i} \bigotimes_{i = 1}^k \ket{0}_{Q_i}
\end{equation}
where each register \(Q_i\) contains \(\left\lceil \log_2(k + 1) \right\rceil\) qudits.  Then apply algorithm \ref{row-form-inversion-algorithm}.  This results in the state
\begin{equation}
  \bigotimes_{i = 1}^k \ket{\pi(i)}_{P_i} \bigotimes_{i = 1}^k \ket{\pi^{-1}(i)}_{Q_i}
\end{equation}
\begin{algorithm}
  \begin{center}
    \begin{algorithmic}[1]
      \For{\(i = 1, \ldots, k\)}
        \For{\(j = 1, \ldots, k\)}
          \State XOR the contents of register \(Q_j\) with \(\ket{i}\) conditional on register \(P_i\) being equal to \(\ket{j}\)
        \EndFor
      \EndFor
    \end{algorithmic}
  \end{center}
  \caption{Computing the inverse of the row representation of \(\pi\)}
  \label{row-form-inversion-algorithm}
\end{algorithm}
Applying algorithm \ref{row-form-inversion-algorithm} again with the roles of the registers \(P_i\) and \(Q_i\) swapped uncomputes the contents of the \(P_i\) registers.  This yields the state
\begin{equation}
  \bigotimes_{i = 1}^k \ket{0}_{P_i} \bigotimes_{i = 1}^k \ket{\pi^{-1}(i)}_{Q_i}
\end{equation}
which results in the row form
\begin{equation}
  \bigotimes_{i = 1}^k \ket{\pi^{-1}(i)}_{Q_i}
\end{equation}
after discarding the registers \(P_i\).  The cost of this procedure is \(O(k^2 \log k)\) basic operations.

\section{Quantum algorithms for state symmetrization}
In this section, the problem of quantum state symmetrization which is closely related to graph isomorphism will be studied.  Let \(\ket{\psi_1}, \ldots, \ket{\psi_n}\) be a sequence of orthonormal states and let \(G\) be a subgroup of \(S_n\).  The problem is to prepare the state \(\frac{1}{\sqrt{\abs{G}}} \sum_{\pi \in G} \bigotimes_{i = 1}^n \ket{\psi_{\pi^{-1}(i)}}\).  Consider a graph \(G\) with adjacency matrix \(A\).  Suppose that it was possible to efficiently prepare the state \(\ket{G} = \sqrt{\frac{\abs{Aut(G)}}{n!}} \sum_{\pi \in S_n / Aut(G)} \ket{A^\pi}\) where \(Aut(G)\) denotes the automorphism group of \(G\) and \(A^\pi\) denotes the adjacency matrix obtained by applying \(\pi\) to \(A\).  Then given two graphs \(G\) and \(H\), one could create the states \(\ket{G}\) and \(\ket{H}\) and these states would be the same if \(G\) and \(H\) were isomorphic and orthogonal otherwise.  The swap test \cite{buhrman2001a} could then be used to determine if \(G\) and \(H\) were isomorphic \cite{aharonov2003a}.  The only difference between the state symmetrization problem and the state preparation approach to graph isomorphism is that in the former, symmetrization is performed over a sequence of orthogonal states whereas in the latter symmetrization is performed on the adjacency matrix.  Moreover, the quantum algorithm for tree isomorphism shown in section \ref{tree-isomorphism-algorithm-section} works by repeatedly symmetrizing over \(S_k\).  An algorithm will now be presented for performing state symmetrization over an arbitrary permutation group \(G\) on \(n\) elements.

\subsection{A quantum algorithm for state symmetrization over a permutation group \(G\) on \(n\) elements}
Let \(G\) be an arbitrary permutation group on \(n\) elements for which a generating set \(K\) is known.   It will be shown how to efficiently prepare the state \(\frac{1}{\sqrt{\abs{G}}} \sum_{\pi \in G} \bigotimes_{i = 1}^n \ket{\psi_{\pi^{-1}(i)}}\) where \(\ket{\psi_1}, \ldots, \ket{\psi_n}\) is a sequence of orthonormal states as before.  The algorithm works using a classical group theoretic algorithm as a subroutine.  Consider the tower of subgroups
\begin{equation}
  G^{(0)} \geq G^{(1)} \geq \cdots \geq G^{(n)}
\end{equation}
where \(G^{(i)} = G_{1, \ldots, i}\) with \(G_{1, \ldots, i}\) denoting the pointwise stabilizer of the set \(\lbrace 1, \ldots, i \rbrace\) in \(G\).  Note that \(G^{(n)}\) is the trivial group.  Now let \(U_i\) be a left transversal of \(G^{(i)}\) in \(G^{(i - 1)}\) --- that is, a set containing exactly one representative of each left coset of \(G^{(i)}\) in \(G^{(i - 1)}\).  Then observe that each of the permutations in \(G\) may be uniquely expressed as a product of the form
\begin{equation}
  \label{permutation-product-form}
  g_1 \cdots g_n
\end{equation}
where each \(g_i \in U_i\).  This can be thought of a generalized version of the subgroup tower form utilized in the quantum algorithm for tree isomorphism.  Now, there is a classical algorithm \cite{furst1980a, hoffman1982a} which can find the left transversals \(U_i\) given the generating set \(K\) for \(G\) using \(O(\abs{K} n^2 + n^6)\) operations.  Moreover, it can be shown that the index of \(G^{(i)}\) in \(G^{(i - 1)}\) is at most \(n - i + 1\).  This allows symmetrization over \(G\) to be performed efficiently as follows.  First, create the state
\begin{align}
  \frac{1}{\sqrt{\abs{G}}} \sum_{\pi \in G} \ket{\pi} = & \bigotimes_{i = 1}^n \frac{1}{\sqrt{\abs{U_i}}} \sum_{g_i \in U_i} \ket{g_i}_{G_i} \\
                                                   = & \frac{1}{\sqrt{\abs{G}}} \sum_{\begin{array}{c}\pi = g_1 \cdots g_n \\ g_i \in U_i\end{array}} \bigotimes_{i = 1}^n \ket{g_i}_{G_i}
\end{align}
where \(\ket{\pi}\) denotes the permutation \(\pi\) represented as a product of the form shown in equation \pref{permutation-product-form}.  Note that this can be done efficiently since it is easy to prepare each of the states \(\frac{1}{\sqrt{\abs{U_i}}} \sum_{g_i \in U_i} \ket{g_i}\) and the desired state is simply their tensor product.  The representation used for each permutation \(\pi\) will now be changed to row form.  Add a new set of registers yielding the state
\begin{equation}
  \frac{1}{\sqrt{\abs{G}}} \sum_{\begin{array}{c}\pi = g_1 \cdots g_n \\ g_i \in U_i\end{array}} \bigotimes_{i = 1}^n \ket{g_i}_{G_i} \bigotimes_{i = 1}^n \ket{i}_{P_i}
\end{equation}
where each register \(P_i\) is initialized to \(\ket{i}\) and uses \(\left\lceil \log_2(n + 1) \right\rceil\) qubits.  The next step is to compute the row form of each permutation \(g_1 \cdots g_n\) in the registers \(P_i\).  This is done using algorithm \ref{row-form-computation-algorithm} and results in the state
\begin{equation}
  \frac{1}{\sqrt{\abs{G}}} \sum_{\begin{array}{c}\pi = g_1 \cdots g_n \\ g_i \in U_i\end{array}} \bigotimes_{i = 1}^n \ket{g_i}_{G_i} \bigotimes_{i = 1}^n \ket{\pi(i)}_{P_i}
\end{equation}
\begin{algorithm}
  \begin{center}
    \begin{algorithmic}[1]
      \For{\(i = 1, \ldots, n\)}
        \For{\(g_i \in U_i\)}
          \State Multiply the permutation represented in row form by the registers \(P_1, \ldots, P_n\) on the right by \(g_i\) conditional on the contents of register \(G_i\) being \(g_i\)
        \EndFor
      \EndFor
    \end{algorithmic}
  \end{center}
  \caption{Computing the row form of each permutation}
  \label{row-form-computation-algorithm}
\end{algorithm}
It is now necessary to uncompute the contents of the \(G_i\) registers.  Now, given the left transversals \(U_i\), there is an efficient classical algorithm \cite{hoffman1982a} which takes a permutation \(\pi \in G\) and computes values \(g_i \in U_i\) such that \(\pi = g_1 \cdots g_n\).  The left transversals \(U_i\) can be regarded as fixed so that the only input of the algorithm is the permutation \(\pi\).  The algorithm therefore takes an input of fixed size so it can be converted to a classical circuit of polynomial size which computes the values \(g_i\).  This circuit can therefore be utilized to uncompute the contents of each \(G_i\) register which yields the state
\begin{equation}
  \frac{1}{\sqrt{\abs{G}}} \sum_{\begin{array}{c}\pi = g_1 \cdots g_n \\ g_i \in U_i\end{array}} \bigotimes_{i = 1}^n \ket{\pi(i)}_{P_i}
\end{equation}
after the \(G_i\) registers are discarded.  This can be rewritten as
\begin{equation}
  \frac{1}{\sqrt{\abs{G}}} \sum_{\pi \in G} \bigotimes_{i = 1}^n \ket{\pi(i)}_{P_i}
\end{equation}
Now, append the orthonormal states \(\ket{\psi_1}, \ldots, \ket{\psi_n}\).  This results in the state
\begin{equation}
  \frac{1}{\sqrt{\abs{G}}} \sum_{\pi \in G} \bigotimes_{i = 1}^n \ket{\pi(i)}_{P_i} \bigotimes_{i = 1}^n \ket{\psi_i}_{B_i}
\end{equation}
Applying a permutation to the order of the \(B_i\) registers conditional on the permutation represented by the registers \(P_i\) results in
\begin{equation}
  \frac{1}{\sqrt{\abs{G}}} \sum_{\pi \in G} \bigotimes_{i = 1}^n \ket{\pi(i)}_{P_i} \bigotimes_{i = 1}^n \ket{\psi_{\pi^{-1}(i)}}_{B_i}
\end{equation}
Next, use algorithm \ref{row-form-inversion-algorithm} to invert each permutation \(\pi\) represented by the registers \(P_i\).  This yields the state
\begin{equation}
  \frac{1}{\sqrt{\abs{G}}} \sum_{\pi \in G} \bigotimes_{i = 1}^n \ket{\pi^{-1}(i)}_{P_i} \bigotimes_{i = 1}^n \ket{\psi_{\pi^{-1}(i)}}_{B_i}
\end{equation}
Algorithm \ref{state-symmetrization-row-form-uncomputation-algorithm} is then applied to uncompute the contents of the \(P_i\) registers.  The result is
\begin{algorithm}
  \begin{center}
    \begin{algorithmic}[1]
      \For{\(i = 1, \ldots, n\)}
        \For{\(j = 1, \ldots, n\)}
          \State XOR the contents of register \(P_j\) with \(i\) conditional on the contents of register \(B_j\) being \(\ket{\psi_i}\)
        \EndFor
      \EndFor
    \end{algorithmic}
  \end{center}
  \caption{Uncomputing \(P_i\) for all \(1 \leq i \leq n\)}
  \label{state-symmetrization-row-form-uncomputation-algorithm}
\end{algorithm}
\begin{equation}
  \frac{1}{\sqrt{\abs{G}}} \sum_{\pi \in G} \bigotimes_{i = 1}^n \ket{0}_{P_i} \bigotimes_{i = 1}^n \ket{\psi_{\pi^{-1}(i)}}_{B_i}
\end{equation}
which yields the desired state
\begin{equation}
  \frac{1}{\sqrt{\abs{G}}} \sum_{\pi \in G} \bigotimes_{i = 1}^n \ket{\psi_{\pi^{-1}(i)}}_{B_i}
\end{equation}
after the \(P_i\) registers are discarded.

\section{Conclusion}
This paper presented a new quantum algorithm which is capable of deciding tree isomorphism in polynomial time; although there is no difficulty in deciding tree isomorphism in polynomial time on classical computers, this quantum algorithm relies on new techniques which may be useful for more general quantum algorithms.  The state symmetrization problem was also discussed and a quantum algorithm was shown which can be used to symmetrize a sequence of \(n\) orthonormal states over any subgroup of \(S_n\).

\section{Acknowledgments}
I thank Professors Dave Bacon and Aram Harrow for feedback and many useful discussions.  In particular, Professor Bacon suggested the use of the concept I call ``subgroup tower form'' which was important in the development of the quantum algorithms for state symmetrization.  Professor Harrow suggested the idea of applying \(U_i^\dagger\) inside the tree isomorphism algorithm.  This allows operations to be controlled by isomorphism of subtrees which allows permutations to be uncomputed.  This work was supported by the NSF under grant CCF-0916400.

\bibliographystyle{unsrt}
\bibliography{$HOME/LaTeX/computer-science-references,$HOME/LaTeX/quantum-computing-references}

\end{document}